\documentclass[aps,prd,amsmath,amssymb,superscriptaddress,twocolumn,nofootinbib,preprintnumbers,showpacs]{revtex4}
\usepackage{graphicx}
\usepackage{hyperref}
\usepackage{color}
\usepackage{verbatim}
\usepackage{soul}
\usepackage[normalem]{ulem}

\begin{document}

\title{Dispersive analysis of $\omega/\phi \rightarrow 3\pi,\,\pi \gamma^*$}

\author{I.V. Danilkin}\email{danilkin@jlab.org}
\affiliation{Center for Theoretical and Computational Physics, Thomas Jefferson National Accelerator Facility, Newport News, VA 23606}

\author{C. Fern\'andez-Ram\'{\i}rez}
\affiliation{Center for Theoretical and Computational Physics, Thomas Jefferson National Accelerator Facility, Newport News, Virginia 23606, USA}

\author{P. Guo}
\affiliation{Center for Exploration of Energy and Matter, Indiana University, Bloomington, Indiana 47403}
\affiliation{Physics Department, Indiana University, Bloomington, Indiana 47405, USA}

\author{V. Mathieu}
\affiliation{Center for Exploration of Energy and Matter, Indiana University, Bloomington, Indiana 47403}
\affiliation{Physics Department, Indiana University, Bloomington, Indiana 47405, USA}

\author{D. Schott}
\affiliation{Department of Physics, The George Washington University, Washington, DC 20052, USA}

\author{M. Shi}
\affiliation{Center for Theoretical and Computational Physics, Thomas Jefferson National Accelerator Facility, Newport News, VA 23606}
\affiliation{Department of Physics, Peking University, Beijing 100871, China}

\author{A. P. Szczepaniak}
\affiliation{Center for Theoretical and Computational Physics, Thomas Jefferson National Accelerator Facility, Newport News, VA 23606}
\affiliation{Center for Exploration of Energy and Matter, Indiana University, Bloomington, IN 47403}
\affiliation{Physics Department, Indiana University, Bloomington, IN 47405}

\collaboration{Joint Physics Analysis Center}

\begin{abstract}
The decays $\omega/\phi \rightarrow 3\pi$ are considered in the dispersive framework that is
based on the isobar decomposition and subenergy unitarity. The inelastic contributions are parametrized by
the power series in a suitably chosen conformal variable that properly accounts for the analytic properties of the amplitude. The Dalitz plot distributions and integrated decay widths are presented. Our results indicate that the final-state interactions may be sizable. As a further application of the formalism we also compute the electromagnetic transition form factors of $\omega/\phi \rightarrow \pi^0\gamma^*$.
\end{abstract}

\date{\today}

\preprint{JLAB-THY-14-1960}

\maketitle
\section{Introduction}

Three-particle production plays an important role in hadron physics. In the past, analysis of the three-pion spectrum led to the discovery of several prominent meson resonances~\cite{PDG-2012}. With the high-precision data already available --- for example, from the COMPASS Collaboration \cite{Adolph:2014uba} and expected from Jefferson Lab \cite{Battaglieri:2010zza} --- in the near future it will be possible to further resolve the three-pion spectrum and identify new resonances that do not necessarily fit the quark-model template. Indeed, in the charmonium spectrum several candidates for nonquark model resonances have recently been reported~\cite{Aaij:2014jqa, Swanson:2006st}. Several of these were observed in decays to three-particle final states. A proper description of interactions in the three-particle system is also required to advance lattice gauge computations of scattering amplitudes~\cite{Polejaeva:2012ut, Briceno:2012rv, Dudek:2014qha, Guo:2013qla}.

Because of large production yields, hadron systems are  also an important laboratory for studies of weak interactions, symmetry tests, and searches for physics beyond the Standard Model~\cite{Celis:2013xja, Daub:2012mu}. Sensitivity to weak interactions demands high precision in the determination of hadronic amplitudes. Near threshold there are first principle constraints that can help in this process. These low-energy constraints include, for example, chiral symmetry, partial wave and effective range expansions, and unitarity. In general, however, it is impossible to construct a single analytical function that describes a reaction amplitude in the entire range of kinematical variables and satisfies all of the constraints imposed by the relativistic $S$-matrix theory. Nevertheless, analyticity is a powerful constraint that enables one to connect different regions of the spectrum {\it e.g.} constrain resonance parameters by the behavior of the amplitude elsewhere, including both the near threshold and high-mass regions.

In this paper we focus on the analysis of three pion production at low energies in particular from decays of the light-vector, isoscalar mesons, the $\omega$ and the $\phi$. At low energies, chiral perturbation theory ($\chi$PT) serves as a powerful constraint on amplitudes involving the light pseudo-scalar mesons \cite{Weinberg:1967tq, Gasser:1983yg}. $\chi$PT has been applied to the three pion production from the $\eta$ decays \cite{Gasser:1984pr, Bijnens:2007pr}. In the case of $\omega/\phi\rightarrow 3\pi$, $\chi PT$ can be extended by including light vector mesons as additional degrees of freedom \cite{Klingl:1996by, Leupold:2008bp, Terschlusen:2012xw,Terschlusen:2013iqa}. In a perturbative study, germane to an effective field theory, unitarity is only satisfied order-by-order in the loop expansion. On the other hand, from the perspective of the $S$-matrix theory, unitarity is the key feature that constrains singularities of the reaction amplitude and therefore the amplitude itself. For this reason there has been a lot of interest in application of dispersion relations to the low energy production of pseudoscalar mesons ~\cite{GarciaMartin:2011cn, Ditsche:2012fv, Dai:2014zta, Gasparyan:2010xz, Hanhart:2013vba, Moussallam:2013una}.

\begin{figure*}[t]
\centering
\includegraphics[keepaspectratio,width=0.9\textwidth]{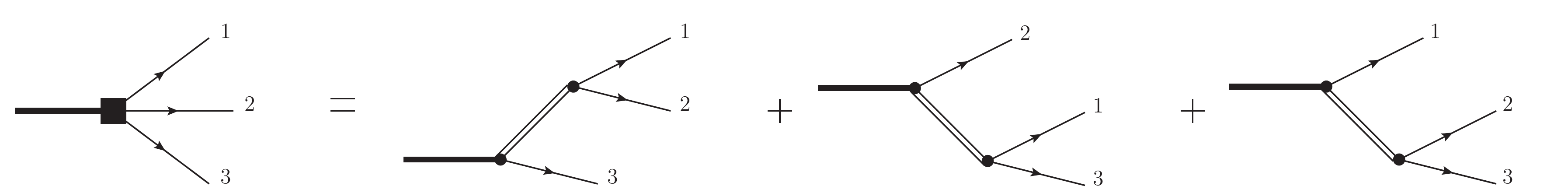}
\caption{\label{Fig:Isobar}Isobar decomposition.}
\end{figure*}

\begin{figure}[t]
\centering
\includegraphics[keepaspectratio,width=0.4\textwidth]{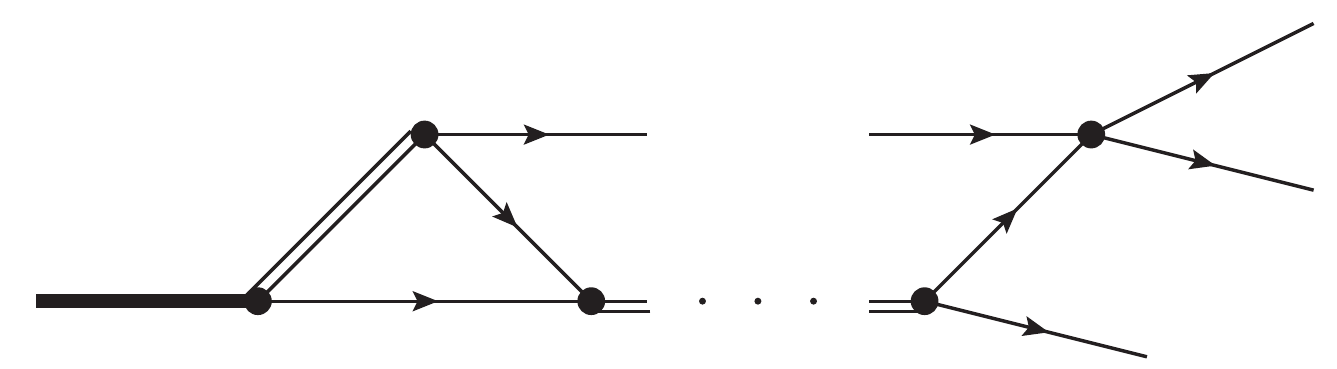}
\caption{\label{Fig:Isobar_Correction}Crossed channel rescattering effects.}
\end{figure}

In the past, dispersive methods have been used in the description of relativistic three body decays \cite{PhysRev.119.1115,PhysRev.132.2703,PhysRev.170.1294,Pasquier:1969dt}. For example, the decay $\eta\rightarrow 3\pi$ \cite{Kambor:1995yc,Anisovich:1996tx,Schneider:2010hs,Colangelo:2009db,Lanz:2013ku,Kampf:2011wr}, is of interest because it is sensitive to isospin breaking, which in QCD originates from the mass difference between the up and down quarks. A dispersive analysis of $\omega$ decay was performed in  \cite{Aitchison:1977ej} and more recently in \cite{Niecknig:2012sj}. It is of interest because it sheds light on the vector mesons dominance and the interplay between the QCD dynamics, which is believed to be responsible for the vector meson formation and its decay characteristics restricted by unitarity and long-range interactions.

In relativistic $S$-matrix theory a function connecting four external particles describes the reaction amplitudes of all processes related by crossing, {\it i.e.} the three  $2\to 2$ scattering channels and, if kinematically allowed, a decay channel $1 \to 3$. Therefore, unitary constraints ought to be considered in all physical channels connected by the same analytical function. With the emphasis on unitarity, the natural starting point for amplitude construction is the partial wave expansion. At low energies, it is expected that only low partial waves are significant and therefore the infinite partial waves series can be truncated to a finite sum. We refer to such an approximation as the isobar model \cite{Herndon:1973yn}.  The diagrams representing a truncated partial waves series (also known as the isobar decomposition) are shown in Fig.\ref{Fig:Isobar}.

The implementation of unitarity on a truncated set of partial waves leads to the so called  Khuri-Treiman (KT) equations \cite{PhysRev.119.1115, PhysRev.132.2703, Aitchison:1976nk}. In the KT framework elastic unitarity in the three crossed channels is used to determine the discontinuity of partial waves which are then reconstructed using a Cauchy dispersion relation. Consequently additional diagrams contribute to the amplitude, see Fig. \ref{Fig:Isobar_Correction}. Since, as discussed above, the model truncates the number of partial waves, it is intrinsically restricted to low energies. In other words the high-energy behavior in the KT framework is arbitrary.  Mathematically, this translates into an arbitrariness in choosing the boundary condition for the solution of an integral equation, which follows from the dispersion relation. It is therefore more appropriate to consider the KT framework as a set of constraints on partial wave equations. Furthermore, above the threshold of the production of inelastic channels the KT amplitudes will couple to other open channels. Any scheme that tries to reduce the sensitivity of the elastic KT equations to the high-energy contributions in dispersion integrals should therefore take into account the change in the analytical properties of the partial wave amplitudes above the inelastic open channels. A novel implementation of this feature within the KT framework  is the main new ingredient of the approach presented in this paper.

In previous works, in order to suppress sensitivity to the unconstrained high-energy region, subtracted dispersion relations were used \cite{Niecknig:2012sj, Colangelo:2009db, Lanz:2013ku}. Moreover, KT equations depend on the elastic $2\to 2$ scattering amplitudes. The $\pi\pi \to \pi\pi$ amplitudes needed for an analysis of $\omega/\phi$ decays have been studied in Ref.~\cite{GarciaMartin:2011cn}. These studies constrained the amplitudes only up to certain center of mass energy (somewhat above the $\bar{K}K$ threshold), and this adds further uncertainty to the KT framework. For example, in previous analyses of the vector meson decays the $\pi\pi$ phase shift was  extended beyond the elastic region with a specific model \cite{Niecknig:2012sj}. In this paper we present an alternative to the subtraction procedure, which not only suppresses the high-energy contributions to the dispersive integrals, but also takes into account the change in the analytical properties induced by the opening of inelastic channels. Specifically, we split the dispersive integral into elastic and inelastic parts, and parametrize the latter in terms of an appropriately chosen conformal variable.

The paper is organized as follows. In the next section we summarize the derivation and main features of the KT framework as applied to the vector meson decays. The discontinuity relation and the role that inelastic effects play in choosing a suitable solution of the dispersive relation are discussed in Sections~\ref{Sec:disc} and \ref{Sec:sol}. The numerical analysis of $\omega/\phi \rightarrow 3\pi$ is presented in Section~\ref{Sec:o}. In Section~\ref{Sec:em}  we consider the electromagnetic (EM) transition form factors of $\omega/\phi \rightarrow \pi^0\gamma^*$ as a further application of our formalism. A summary and outlook are presented in Section~\ref{Sec:Conclusion}.

\section{Partial wave or Isobar decomposition}
\label{Sec:1}
The matrix element for the three pion decay of a vector particle is given in terms of a helicity amplitude $H^{abc}_\lambda$, 
\begin{multline}
\langle \pi^a(p_1) \pi^b(p_2) \pi^c(p_3)\,|\,T\,|\,V(p_V,\lambda)\rangle=\\
=(2\pi)^4\,\delta(p_V-p_1-p_2-p_3)H^{abc}_{\lambda}.
\end{multline}
Here $p_V$ and $\lambda$ are the momentum and helicity of the vector particle, $V=\omega/\phi$ in our case, 
$p_1,p_2$, and $p_3$ are the momenta of outgoing pions  with $a,b,$ and $c$ denoted by their Cartesian isospin indices. The Lorentz-invariant Mandelstam variables are defined by $s=(p_V-p_3)^2$, $t=(p_V-p_1)^2$, and $u=(p_V-p_2)^2$, and satisfy the relation
\begin{equation}
s+t+u=M^2+3\,m_\pi^2\,.
\end{equation}
The helicity amplitude $H^{abc}_\lambda$ can be expressed in terms of a single scalar function of the Mandelstam variables, since Lorentz and parity invariance imply that, 
\begin{equation}\label{Eq:covariant_form}
H^{abc}_{\lambda }= i \, \epsilon _{\mu \nu \alpha \beta }\,\epsilon ^{\mu}(p_V,\lambda )\,
p_1^{\nu }\,p_2^{\alpha }\,p_3^{\beta }\,\frac{P^1_{abc}}{\sqrt{2}}\,F(s,t,u)\,,
\end{equation}
where $P^1_{abc} = -i\,\epsilon_{abc}/\sqrt{2}\,$ is the isospin factor corresponding to the coupling of three isospin-1 pions to a state with total isospin-0. The invariant amplitude $F(s,t,u)$ satisfies Mandelstam analyticity \cite{Mandelstam:1958xc,Mandelstam:1959bc} which postulates that it is an analytic function everywhere except for cuts required by unitarity. The scalar function $F(s,t,u)$ is free from kinematical singularities~\cite{Lutz:2011xc,Heo:2014cja}. The latter appear in the covariant factor in front of $F(s,t,u)$ in Eq.~(\ref{Eq:covariant_form}). Crossing symmetry implies that the function $F(s,t,u)$ describes the decay $V \to 3\pi$ and also the three $V \pi \to 2\pi$ scattering channels. Since we are interested in a partial wave decomposition it is necessary to consider the helicity amplitude, $H_\lambda^{abc}$ first. In the physical region of $s$-channel scattering, $V(p_V,\lambda)\pi^c(p_{\bar 3}) \to \pi^a(p_1) \pi^b(p_2)$, the Mandelstam variable $s=(p_V+p_{\bar 3})^2 =(p_V-p_3)^2$ corresponds to the square of the center of mass energy and $t = (p_V - p_1)^2$ is related to the cosine of the $s$-channel scattering angle by 
\begin{equation} \label{zs} 
z_s = \cos\theta_s = \frac{t -u}{4\,p(s)\,q(s)}  \equiv \frac{t -u}{k(s)} 
\end{equation} 
where
\begin{gather}
q(s)=\frac{\lambda^{1/2}(m_\pi^2,m_\pi^2,s)}{2\sqrt{s}}\,,\quad p(s)=\frac{\lambda^{1/2}(M^2,m_\pi^2,s)}{2\sqrt{s}}
\end{gather}
are the magnitude of the relative momentum between the outing pions in the $s$-channel center of mass frame and the magnitude of the incoming pion's momentum in the same frame, respectively. $\lambda(x,y,z)=x^2+y^2+z^2-2\,(xy+yz+xz)$ is the K\"all\'en triangle function. The $s$-channel partial wave decomposition is  given by \cite{Jacob:1959at}
\begin{eqnarray}\label{Eq:p.w._expansion}
H^{abc}_{\lambda}=\frac{P^1_{abc}}{\sqrt{2}}\, \sum_{J=1,3,\dots}\, (2J+1)\,d_{\lambda 0}^J(\theta_s)\,f^J_\lambda(s) 
\end{eqnarray}
were $d_{\lambda 0}^J(\theta_s)$ are the Wigner d-functions and we choose the $x-z$ plane as the reaction plane. Due to Bose symmetry the sum over partial waves is restricted to odd values of $J$ and parity conservation implies that $f^J_0(s)=0$ and  $f^J_{+1}(s) = - f^J_{-1}(s) \equiv f_J(s)$. Therefore there is only one independent helicity amplitude, which is consistent with there being a single scalar function, $F(s,t,u)$ describing the strong coupling between an isoscalar vector and three pions. The relation between $H^{abc}_\lambda$  and $F(s,t,u)$ in Eq.~(\ref{Eq:covariant_form}) enables the determination of the kinematical singularities of the partial wave amplitudes $f_J(s)$. Expressing the Wigner $d$-functions in terms of Legendre polynomials (with a prime denoting a derivative) 
\begin{equation}\label{Eq:d-funciton}
d^J_{10}(\theta)=-\frac{\sin\theta}{\sqrt{J(J+1)}}\,P'_J(\cos\theta)\,.
\end{equation}
and defining the reduced partial waves $F_J(s)$ by 
\begin{equation}\label{Eq:new_ampl}
F_{J}(s) \equiv \frac{\sqrt{2}}{\sqrt{s}\,p(s)\,q(s)} \frac{2J+1}{\sqrt{J(J+1)}}  \frac{f_{J}(s)}{(p(s)\,q(s))^{J-1}} 
\end{equation} 
the series in Eq.~(\ref{Eq:p.w._expansion})  becomes, 
\begin{eqnarray}\label{Eq:p.w._expansion-2}
H^{abc}_{+}=-P^1_{abc}\frac{\sqrt{\phi}}{4} \sum_{J=1,3,\,\dots} (p(s)q(s))^{J-1}P'_J(z_s)\, F_J(s)
\end{eqnarray}
where $\phi$ is the  Lorentz-invariant Kibble function 
\begin{eqnarray}
\phi&=&(2\sqrt{s}\sin\theta\,p(s)\,q(s))^2\nonumber\\
&=& s\,t\,u-m_\pi^2\left(M^2-m_\pi^2\right)^2.\label{Eq:useful_relations}
\end{eqnarray}
Finally, by comparing Eq.~(\ref{Eq:p.w._expansion-2}) with Eq.~(\ref{Eq:covariant_form}) one finds the desired relation between the scalar amplitude  $F(s,t,u)$ and the reduced partial wave amplitudes $F_J(s)$,
\begin{equation}\label{Eq:p.w._dec3}
F(s,t,u)=\sum_{J=1,3,\dots} (p(s)\,q(s))^{J-1} P'_J(z_s)\,F_{J}(s)\,.
\end{equation}
The sum over partial waves runs over odd values of $J$ and the derivative of the Legendre polynomial is an even polynomial in $z_s$ of order $(J-1)$. Therefore the product of the factors in front of $F_J(s)$ in Eq.~(\ref{Eq:p.w._dec3}) is a polynomial in the $s,t$ and $u$ variables and it is therefore free from kinematical singularities. Since $F(s,t,u)$ has only dynamical singularities this implies that the reduced partial waves must also have only the dynamical singularities, and therefore they can be expressed in terms  of  discontinuities across unitary cuts. We note that the decomposition (\ref{Eq:p.w._dec3}) is different from that in Eq.(6) of \cite{Niecknig:2012sj}, where only the p-wave amplitude had its kinematical singularities removed.

We emphasize that in Eq.~(\ref{Eq:p.w._dec3}) the sum extends to infinity. The sum converges in the $s$-channel physical region and it is to be analytically continued to obtain amplitudes in the physical regions of the other two scattering channels or the decay channel. Since in Eq.~(\ref{Eq:p.w._dec3}) each term in the sum is a polynomial in $t$ and $u$, singularities of $F(s,t,u)$ in these variables demanded by the $t$ or $u$ channel unitarity can only emerge from the infinite number of terms in the  sum. The isobar approximation amounts to truncating the partial wave series at a finite value of $J = J_{max}$. In order to retain dynamical singularities of $F(s,t,u)$ in all three variables, in the isobar model, the scalar amplitude is approximated by a linear combination of truncated partial wave series in the three channels simultaneously\footnote{In principle, the isobar decomposition should be written for the full amplitude $H_\lambda^{abc}$, but since the product of the isospin and kinematic factors in Eq.~(\ref{Eq:covariant_form})  is symmetric under permutation of pions we only need to symmetrize $F(s,t,u)$.}, which yields, 
\begin{eqnarray}\label{Eq:covariant_form_2}
F(s,t,u) && =  \sum_{J=1,3,\dots}^{J_{max}} (p(s) q(s))^{J-1} P'_J(z_s)\,F_{J}(s) \nonumber \\
&&  + (s \to t) + (s\to u) 
\end{eqnarray} 
where because of the Bose symmetry, the partial waves in each channel are given by the same function $F_J(x)$ with  $x=s,t,u$.  The $t$ and $u$ channel scattering angles are given by 
\begin{eqnarray} 
& & z_t = \cos\theta_t = \frac{s -u}{4\,p(t)\,q(t)} \nonumber \\
& & z_u = \cos\theta_u = \frac{t -s}{4\,p(u)\,q(u)},
\end{eqnarray} 
respectively. 
The isobar model ansatz of (\ref{Eq:covariant_form_2}) satisfies crossing symmetry and the single-variable dispersion relation. The value of $J_{max}$ should be determined by comparing with the experimental data. Note that for large $J_{max}$ the model becomes unreliable since, as discussed above, the truncation of a partial wave series introduces an incorrect dependence on the cross-channel energy variable, {\it i.e.} at $J=J_{max}$ the $s$-channel series behaves as $(t-u)^{J_{max}-1}$. We also note, that in the general representation of the full amplitude as a sum of functions that are singular in one variable at the time follows from the Mandelstam double spectral representation (the opposite is not true).  To prove this, one has to assume two-body unitarity and truncate the partial wave expansion of the amplitude. For $\pi\pi$ scattering a similar decomposition \cite{Stern:1993rg,Knecht:1995tr,Zdrahal:2008bd} was shown to be true up to next-to-next-to-leading order in $\chi$PT.
   
\begin{figure*}[t]
\center{\includegraphics[keepaspectratio,width=0.48\textwidth]{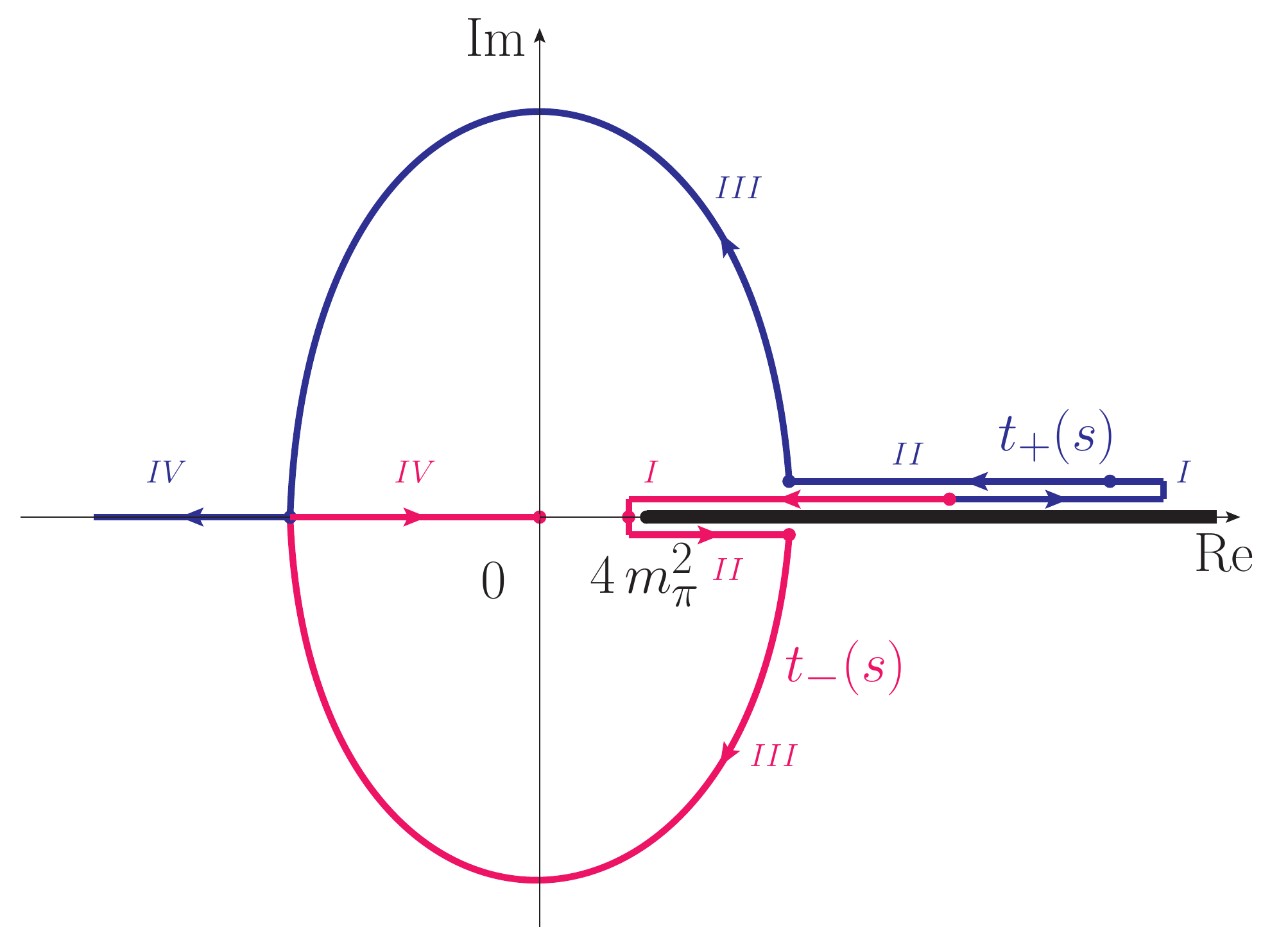}
\includegraphics[keepaspectratio,width=0.48\textwidth]{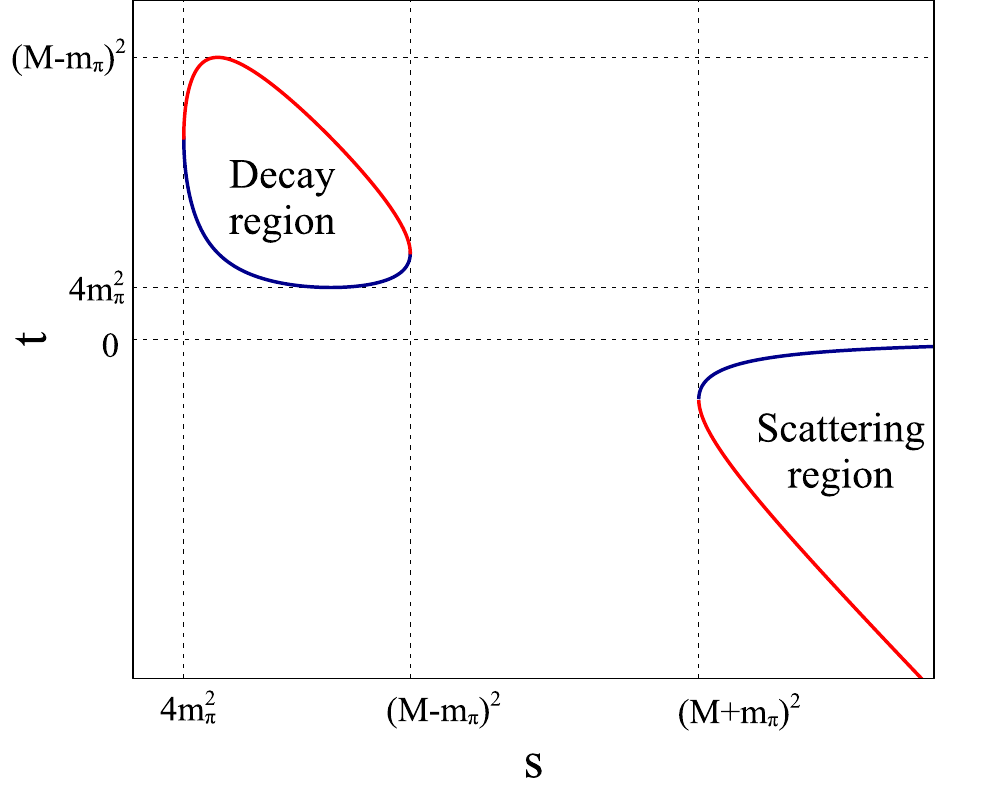}}
\caption{\label{Fig:Contour}Left: Integration contour. The analytical continuation from the scattering region to the decay region is made by adding a positive infinitesimal imaginary part to the vector meson mass: $M^2\to M^2+i\epsilon$ \cite{PhysRev.132.2703}. Right: Part of the Mandelstam plane, where the decay and $s$-channel scattering regions are shown.}
\end{figure*}

The isobar model helicity amplitude 
can therefore be written as 
\begin{multline}\label{Eq:isobar_dec2}
H^{abc}_{\lambda }= i\, \epsilon _{\mu \nu \alpha \beta }\,\epsilon ^{\mu}(p_V,\lambda)\,
p_+^{\nu }\,p_-^{\alpha }\,p_0^{\beta }\,\frac{P^1_{abc}}{\sqrt{2}}\sum_{J=1,3,\dots}^{J_{max}}\\ \Big(\tilde P_l(z_s)\,F_{J}(s)+\tilde 
P_J(z_t)\,F_{J}(t)+\tilde P_J(z_u)\,F_{J}(u)\Big)\,,
\end{multline}
where we defined $\tilde P_J(z_x) \equiv (p(x)q(x))^{J-1} P'_J(z_x)$ with $x=s,t,u$. 
This expression coincides with the expression used in Ref. \cite{Niecknig:2012sj} for $J=1$.
The three diagrams in Fig. \ref{Fig:Isobar} represent individual partial waves in the $s$, $t$, and $u$ channels.

\section{Two particle discontinuity relation}
\label{Sec:disc}

We constrain the reduced partial waves by imposing elastic unitarity. Because of Bose symmetry it is sufficient to consider  the constraint in a single, {\it e.g.} $s$-channel. At fixed $t$ and $s$ in the $s$-channel physical region, the discontinuity
$$\text{Disc}\,F(s,t,u) =\frac{1}{2\,i}\big( F(s+ i\epsilon,t,u) - F(s-i\epsilon,t,u)\big)$$
is computed by taking the partial wave projection of the unitarity relation for the helicity amplitude, 
\begin{eqnarray}\label{Eq:Unitarity_1}
&&\text{Disc}\,H^{a b c}_{\lambda}\left(p_V p_{\bar 3} \rightarrow  p_1p_2\right)\nonumber \\
&&\qquad\qquad=\frac{1}{4}\int d \Phi\,t^{*aba'b'}(q_1'q_2'\rightarrow p_1p_2)\\
&&\qquad\qquad\quad\times H_{\lambda}^{a' b' c}(p_Vp_{\bar 3} \rightarrow q_1'q_2')\nonumber
\end{eqnarray}
where $t^{aba'b'}$ is the isospin-1 pion-pion scattering amplitude. The integral extends over the two-body $\pi\pi$ phase space. The partial wave expansion for the $\pi\pi \to \pi\pi$ scattering amplitude is given by, 
\begin{gather} \label{pipi} 
t^{aba'b'}=32\pi\sum_{J} (2J+1)\,P_{J}(\cos\theta)\,P^{1}_{aba'b'}\,t_{J}(s)
\end{gather}
where the isospin-1 projection operator is, 
\begin{gather}
P^{1}_{aba'b'}=\frac{1}{2}\,(\delta_{aa'}\delta_{bb'}-\delta_{ab'}\delta_{ba'})\,.
\end{gather}
We consider only the elastic two-particle unitarity since the KT model is restricted to low energies.

Using Eqs.~(\ref{Eq:Unitarity_1}) and (\ref{pipi}) one obtains the expression for the discontinuity of the $s$-channel partial waves
\begin{gather}
\text{Disc}\,F_J(s)=\rho (s)\,t_J^*(s)\Big(F_J(s)+2\,\frac{(2J+1)}{J(J+1)}\sum_{J'=1,3,\dots}^{J'_{max}}\nonumber\\
\int_{-1}^1\frac{d z_s}{2}\,\frac{\phi}{4}\,\frac{\tilde{P}_J(z_s)}{s\,(p(s)\,q(s))^{2J}}\,\tilde{P}_{J'}(z_t)\,F_{J'}(t)\Big)\,, \label{Eq:Disc1}\\
\rho (s)=(1-4\,m_{\pi}^2/s)^{1/2}\,.\nonumber
\end{gather}
The first term on the right hand side originates from the $s$-channel partial wave expansion and is therefore diagonal in $J$. The second-term sums the $s$-channel projection of $t$ and $u$-channel partial wave series, isobars, and it mixes $s$-channel partial waves. In the following we consider only the P-waves, {\it i.e.} take $J_{max} = 1$, since in the kinematical region in $s,t,u$ corresponding to the $\omega/\phi$ decays the spin-3 and higher partial waves are expected to be insignificant. Thus in the analysis that follows we use, 
\begin{gather}
\text{Disc}\,F(s)=\rho(s)\,t^*(s)\left(F(s)+\hat{F}(s)\right)\,,\nonumber\\
\hat{F}(s)=3\int _{-1}^{+1}\frac{dz_s}{2}\left(1-z_s^2\right)F(t(s,z_s))\label{Eq:Disc2}
\end{gather}
and, for real $s$,  
\begin{equation}  \label{di} 
F(s) = \frac{1}{\pi} \int_{4\,m_{\pi}^2}^{\infty} ds' \frac{\mbox{Disc} F(s')}{s' - s - i\,\epsilon} 
\end{equation} 
where $F(s)\equiv F_{1}(s)$, $t(s)\equiv t_{1}(s)$,  $\rho(s) = \sqrt{1 - 4m_\pi^2/s}$ is the phase space factor, and the dependence on $t$ under the integral should be expressed in terms of $s$ and $z_s$ using Eq.~(\ref{zs}). The result in Eq.~(\ref{Eq:Disc2}) is consistent with \cite{Aitchison:1976nk, Niecknig:2012sj}. We note, that Eq.~(\ref{Eq:Disc2}) is exact in the elastic region, while for higher energies one has to incorporate inelastic contributions. Also, Eq.~(\ref{Eq:Disc2}) was derived in the physical region of the $s$-channel, which corresponds to $s\geq (M + m_\pi)^2$ and $|z_s|\leq 1$. To obtain $F(s)$ in the decay region $4\,m_\pi^2 \leq s \leq (M-m_\pi)^2$ the right hand side of Eq.~(\ref{Eq:Disc2}) has to be analytically continued in $s$. Analytical  continuation of the direct-channel contribution is well known. However, analytical continuation of the exchange contribution is more difficult and was extensively studied in Refs. \cite{Kambor:1995yc, PhysRev.132.2703}. If the integration over $z_s$ is replaced by integration over $t$, by using Eq.~(\ref{zs}) the exchange contribution becomes 
\begin{equation}
\hat F(s) = \frac{3}{k(s)} \int _{t^-(s)}^{t^+(s)} dt\,(1-z^2_s(s,t)) \,F(t) \label{cont} 
\end{equation} 
with 
\begin{equation} 
t^{\pm}(s)=\frac{M^2+3m_\pi^2-s}{2}\pm\frac{k(s)}{2}\,.
\end{equation} 
In the $s$-channel the limits of integration $t^{\pm}(s)$ lie on the negative real axis (labeled as region $IV$ in Fig.~\ref{Fig:Contour}) and do not overlap with the cut of the integrand extending over the positive real axis above $t=4\,m_\pi^2$. 
As shown in \cite{Kambor:1995yc} analytical continuation to the decay region requires that the integration is deformed to follow a path that does not cross the unitarity cut of $F(t)$ for $t \geq 4\,m_\pi^2$, as shown in Fig.~\ref{Fig:Contour}. It is worth noting that once kinematical singularities have been removed, the $t$ dependence induced by the partial wave projection, the factor $(1-z^2_s(t,s))$ in Eq.~(\ref{cont}) does not have singularities in $t$.  
In the decay region, $\mbox{Disc}\,F(s)$ is a complex function of $s$ with singularities arising from cuts in the barrier factor $k(s)$ {\it c.f.} Eq.~(\ref{zs}). Guided by the analysis of the triangle diagram in perturbation theory, proper determination  of the singularities in $k(s)$ for $s\geq 4\,m_\pi^2$ was given in Ref. \cite{PhysRev.132.2703} and the right analytical structure of $k(s)$ is
\begin{eqnarray}
k(s)&=&\left\lbrace
\begin{array}{ll}
+\kappa(s)\,,&\quad  4\,m_\pi^2\leq s\leq (M-m_\pi)^2\\
\,\,i\,\kappa(s)\,,&\quad (M-m_\pi)^2\leq s\leq (M+m_\pi)^2\\
-\kappa(s)\,,&\quad (M+m_\pi)^2\leq s<+\infty
\end{array}
\right.\nonumber \\
\kappa(s)&=&\frac{1}{s}\,|\lambda(m_\pi^2,m_\pi^2,s)\,\lambda(M^2,m_\pi^2,s)|^{1/2}\,.
\end{eqnarray}

In the next section we discuss solutions of Eq.~(\ref{Eq:Disc2}).

\begin{figure*}[t]
\centering
\includegraphics[keepaspectratio,width=\textwidth]{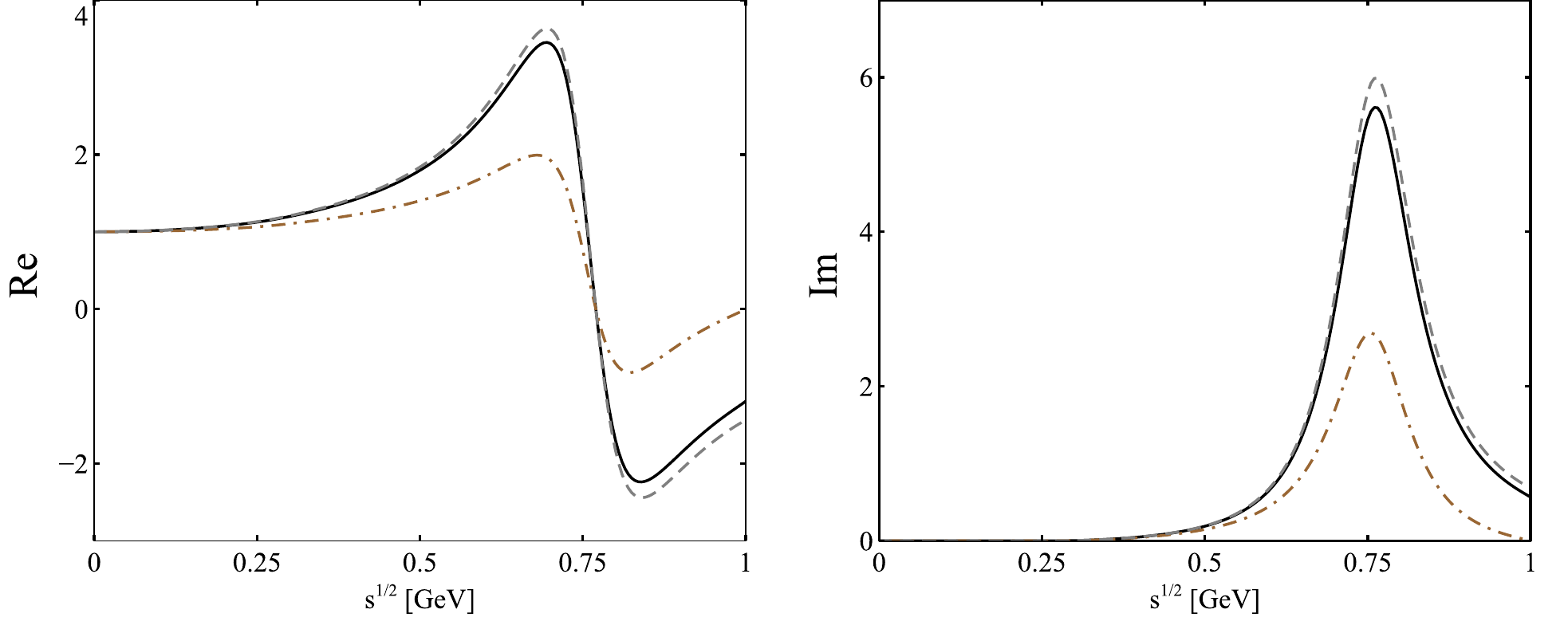}
\caption{\label{Fig:DifferentOmnes}Real and imaginary parts of $\Omega_{el}(s)$ in Eq.~(\ref{Eq:Omnes_el}) (Dot-Dashed), $\Omega'(s)$ in Eq.~(\ref{Eq:Omnes'}) (Dashed) and  $\Omega(s)$ in Eq.~(\ref{Eq:Omnes}) (Solid).}
\end{figure*}

\section{Solution strategies}
\label{Sec:sol}

From the discontinuity (\ref{Eq:Disc2}) one can reconstruct the amplitude using the dispersion relation (\ref{di}). For practical reasons, however, it is useful to represent the amplitude $F(s)$ as a product of two functions
 \begin{equation}\label{Eq:T=OmnesG}
F(s)=\Omega(s)\,G(s)
\end{equation}
where the function $\Omega(s)$ satisfies the following unitarity relation for $s\geq s_\pi =4\,m_\pi^2$
\begin{equation}\label{Eq:DiscOmnes}
\text{Disc}\,\Omega(s)=\rho(s)\,t^*(s)\,\Omega(s)+ \text{inelastic}\,\theta(s>s_i)\,.
\end{equation}
where the first term on the right-hand side represents the elastic contribution. 
The advantage of the representation in Eq.~(\ref{Eq:T=OmnesG}), is that one can absorb the homogeneous part, {\it cf.} first term on the right-hand side in Eq.~(\ref{Eq:Disc2}), into $\Omega(s)$ leaving the contribution from the cross-channel in $G(s)$. Since $F(s)$ and $\Omega(s)$ have only unitary, right-hand cuts, the function $G(s)$ should also have the right-hand cuts. Combining Eqs. (\ref{Eq:Disc2}), (\ref{Eq:T=OmnesG}) and (\ref{Eq:DiscOmnes}) we obtain the following discontinuity relation for $G(s)$
\begin{equation}
\text{Disc}\,G(s)=\frac{\rho(s)\,t^*(s)}{\Omega^*(s)}\,\hat{F}(s)+ \text{inelastic}\,\theta(s>s_i)
\end{equation}
where the last term absorbs inelastic contributions starting with a threshold at $s=s_i$. 
The dispersion relation for $G(s)$  is given by, 
\begin{equation}
G(s)=\int_{s_\pi}^{\infty}\frac{ds'}{\pi}\frac{\text{Disc}\,G(s')}{s'-s}
\end{equation}
where we split the integral into two parts
\begin{equation}
\int_{s_\pi}^{\infty}=\int_{s_\pi}^{s_i}+\int_{s_i}^{\infty}\,.
\end{equation}
The first part is determined entirely by elastic scattering while the second part takes into account inelastic effects.
The inelastic contribution is described by an analytical function on the $s$-plane cut along the real axis above $s=s_i$. 
It is largely unknown, and often parametrized through an expansion in a conformal variable which maps the right-hand cut in the complex $s$-plane onto the unit disk. Such a mapping is known as a convenient representation of functions on a cut plane with the known analytical properties~\cite{Yndurain:2002ud},
\begin{equation}\label{Eq:Sigma}
\Sigma (s)=\sum_{i=0}^{\infty}a_i\,\omega^i(s)
\end{equation}
The variable 
\begin{equation}\label{Eq:Conformal_mapping_function}
\omega(s)=\frac{\sqrt{s_i-s_E}-\sqrt{s_i-s}}{\sqrt{s_i-s_E}+\sqrt{s_i-s}}
\end{equation}
maps the cut plane onto the unit disk. Strictly speaking, the first possible inelastic contributions in the $I=1$, P-wave $\pi\pi\rightarrow \pi\pi$ and $\omega(\phi)\pi\rightarrow \pi\pi$ reactions originates  from the $4\,\pi$ channel. 
However, at low energies they are known to be weak and the parameter $s_i=1$ GeV is identified with the point where inelastic contributions are expected to become relevant. The expansion point $s_E$ should lie below the cut and we define $s_E=0$. The conformal mapping technique was successfully applied in other descriptions of two-to-two amplitudes {\it e.g.} in \cite{Gasparyan:2012km, Danilkin:2011fz, Danilkin:2012ua} it was used to take into account the contributions from the more distant left-hand cuts. However, to the best of our knowledge the conformal mapping technique has never been used before in the context of the KT equations.

With the inelastic contributions parametrized by the function, $\Sigma(s)$, the integral equation for the KT amplitude takes the form of 
\begin{equation}\label{Eq:Integral_eq}
F(s)=\Omega(s)\left(\frac{1}{\pi} \int_{s_\pi}^{s_i}ds' \frac{\rho(s')\,t^*(s')}{\Omega^{*}(s')}\frac{\hat{F}(s')}{s'-s}+\Sigma(s)\right)\,.
\end{equation}
This is an alternative to the standard way which employs subtractions to reduce the sensitivity of the dispersive integral to the high-energy region~\cite{Niecknig:2012sj}. The problem with subtractions is twofold. First, the dispersive integrals, including computation of $\Omega(s)$ run over inelastic regions, while the dispersion relation contains only the elastic contributions. Furthermore, subtracting an analytical function of $s$ does not account for the change in the analytical behavior of the amplitudes due to opening of inelastic channels. 

In Eq.~(\ref{Eq:Integral_eq}) there is no need for subtractions in the dispersive integral since it is restricted to the elastic-region, which is the only part of the right hand cut controlled by elastic unitarity. The unknown, inelastic contributions are parametrized by $\Sigma(s)$, and are to be determined by comparing with the experimental data, or other theoretical approaches that treat inelastic channels explicitly. Moreover, with the dispersive integral restricted to a finite interval over $s$ there are uncontrollably large contributions from higher-partial waves, which otherwise require more and more subtractions. 

\begin{figure*}[t]
\centering
\includegraphics[keepaspectratio,width=\textwidth]{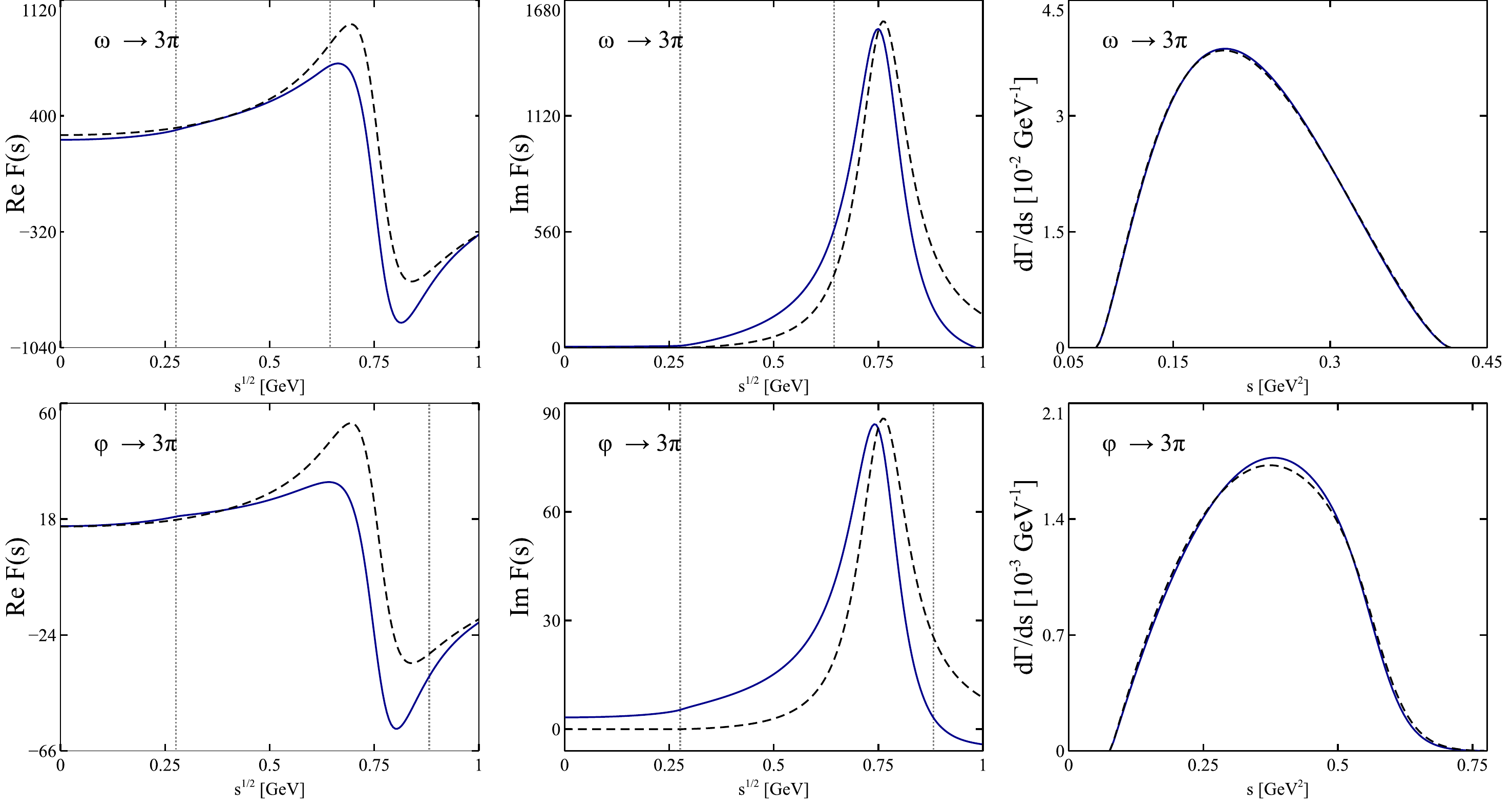}
\caption{\label{Fig:3b_effects}Left and Middle: Solutions of Eq. (\ref{Eq:Integral_eq}) with (solid curves) and without (dashed curves) three body effects. Dotted lines indicate the kinematically allowed region. Right: Single differential decay rate $d\Gamma/ds$.}
\end{figure*}

Besides $F(s)$, the problem with the determination of inelastic contributions also affects the computation of $\Omega(s)$.  The unitarity condition in Eq.~(\ref{Eq:DiscOmnes}) does not determine it above the inelastic threshold $s=s_i$. Therefore, we seek its solution given in terms of the Omn\`es function~\cite{Muskhelishvili-book, Omnes:1958hv} taken only over the elastic region 
\begin{gather}\label{Eq:Omnes_el}
\Omega_{el}(s) \equiv \exp\left(\frac{s}{\pi} \int_{s_\pi}^{s_i} \frac{ds'}{s'}\frac{\delta(s')}{s'-s}\right),
\end{gather}
where $\delta(s)$ is the isospin-1, P-wave $\pi\pi$ phase shift. Because the upper limit is finite, the function $\Omega_{el}(s)$ has a zero at $s=s_i$ \cite{Buettiker:2003pp},
\begin{gather}
\Omega_{el}(s\rightarrow s_i)\sim|s-s_i|^{\alpha(s)}\,,
\end{gather}
where $\alpha(s)=\delta(s)/\pi$. The zero can be removed by defining, 
\begin{gather}\label{Eq:Omnes}
\Omega_{el}(s) \to \Omega(s)=(s_i-s)^{-\alpha(s_i)}\,\,\Omega_{el}(s)\,.
\end{gather}
The factor in front of $\Omega_{el}(s)$ has only the inelastic cut and therefore in the elastic region $\Omega(s)$, just like $\Omega_{el}(s)$ satisfies two body-unitarity relation. 
In Fig. \ref{Fig:DifferentOmnes} we plot $\Omega_{el}(s)$, $\Omega(s)$ and compare them with the standard representation obtained by integrating the elastic phase shift to infinity, 
\begin{equation}\label{Eq:Omnes'}
\Omega'(s)=\exp\left(\frac{s}{\pi} \int_{s_\pi}^{\infty} \frac{ds'}{s'}\frac{\delta(s')}{s'-s}\right)\,
\end{equation}
In $\Omega'$, following \cite{Niecknig:2012sj}, the phase shift is assumed to approach a constant at infinity, $\delta(s\rightarrow \infty)\rightarrow \pi$. This is obtained by  smoothly  matching to the low-energy parametrization form \cite{GarciaMartin:2011cn} at $s=1.3$ GeV. We remark, that  Eq.~(\ref{Eq:Omnes}) is equivalent to Eq.~(\ref{Eq:Omnes'}), when the phase shift is set to a constant value equal to $\delta(s_i)$ for $s\geq s_i$.

Finally, we note that the upper limit in the integral of Eq.~(\ref{Eq:Integral_eq}) also induces an artificial singularity at $s=s_i$. This singularity originates from absorbing any contribution to the dispersive integral over the energy range $s\geq s_i$ into the function $\Sigma(s)$. This singularity is eliminated by adding to the dispersive integral in Eq.~(\ref{Eq:Integral_eq}) the term 
\begin{eqnarray}\label{Eq:Regularization}
-\frac{1}{\pi} \frac{\rho(s_i) \,t^*(s_i)\, \hat F(s_i)}{\Omega^*(s_i)}  
\log\left( \frac{s_i - s}{s_i - s_\pi} \right) 
\end{eqnarray}
which cancels the $\log(s_i - s)$ singularity of the integral as $s \to s_i$. This is the correct  way of regulating this singularity since the added term is a function with an inelastic cut only {\it i.e.}, it can be absorbed into the function $\Sigma(s)$ in Eq.~(\ref{Eq:Integral_eq}). The scale in the denominator of the $\log$ is chosen to have a negligible effect at $s=s_\pi$.

In summary, in Eq. (\ref{Eq:Integral_eq}) $\Omega(s)$ is given on the right hand side of Eq. (\ref{Eq:Omnes}). The expression in (\ref{Eq:Regularization}) is added to the integral inside the parenthesis of (\ref{Eq:Integral_eq}) and $\Sigma(s)$ is represented by (\ref{Eq:Conformal_mapping_function}). As discussed before, these changes do not affect elastic unitarity.

We wish to remark that if we knew the discontinuity relation of the amplitude not only at low energies as in Eq.~(\ref{Eq:Disc2}) but for all energies, then using the analytical properties of the amplitude we could reconstruct the solution everywhere up to a polynomial. However, there are inelastic channel contributions that force us either to introduce the extra subtractions in order to suppress the unknown high energy region or cutoff the integral and parametrize the inelastic contribution by a conformal mapping technique. As discussed earlier in this section, the latter enables the amplitude to retain the analytical properties expected in the presence of inelasticities.

\section{Numerical results}
\subsection{$\omega/\phi \rightarrow 3\pi$}
\label{Sec:o}

We solve the integral equation in Eq.(\ref{Eq:Integral_eq}) by numerical iteration \footnote{Note that the double integral in Eq. (\ref{Eq:Integral_eq}) ($\hat F(s)$ is given by a contour integral over $t$ as shown in Eq. (\ref{Eq:Disc2})) can be inverted using the Pasquier method \cite{PhysRev.170.1294,Aitchison:1978pw}. In this method the order of the $s$ and $t$ integration is reversed, with the latter deformed onto a real axis that can be calculated analytically or numerically only once. This leads to a single-variable integral equation for $\hat F(s)$ with a kernel that depends on the input two-body scattering amplitude. This is an equivalent method to solve the KT equation which has its advantages and disadvantages \cite{Guo:2014vya}.}. The convergence is fast, typically 
after three to four iterations, and no significant variations in the solution are observed. From the amplitude, it is straightforward to compute the Dalitz plot distribution, the partial decay and the total, $3\pi$  decay widths, \cite{PDG-2012}
\begin{eqnarray}
\frac{d^2\Gamma}{ds\,dt}=\frac{1}{(2\pi)^3}\frac{1}{32M^3}\frac{1}{3}P(s,t)\,|F(s,t,u)|^2\,,
\end{eqnarray}
where $P(s,t)=\phi(s,t)/4$ is the kinematic factor discussed in Sec.~\ref{Sec:1}. In the computations of the Dalitz plot that follow, the conformal expansion in Eq.~(\ref{Eq:Sigma}) is truncated at $0^{\mbox{th}}$ order {\it i.e.}  only a constant term is kept and this is the only free parameter of the model. It is fixed to reproduce the measured $3\pi$ decay widths for $\omega$ and $\phi$, which are   $\Gamma_{\omega\rightarrow 3\pi}^{exp}= 7.57\mbox{ MeV}$ and   $\Gamma_{\phi\rightarrow 3\pi}^{exp} = 0.65\mbox{ MeV}$, respectively \cite{PDG-2012}. Since the integral equation is linear in $F(s)$, the one parameter that is fitted is responsible for the overall normalization, while the  Dalitz plot distribution is only affected by higher order terms in $\Sigma(s)$.

In Fig. \ref{Fig:3b_effects} we show the solution of the integral equation (\ref{Eq:Integral_eq}) together with the invariant mass distribution. The significance of the three-body effects, given by the cross-channel terms, is accessed by keeping or eliminating $\hat{F}$ from the discontinuity relation. In either case $\Sigma(s)$ is represented by a constant which is fitted to reproduce the decay width. As can be seen in Fig. \ref{Fig:3b_effects} the effect of the crossed-channels for $\omega\rightarrow 3\pi$ is less significant than for $\phi\rightarrow 3\pi$.  In both cases, the invariant mass distributions are quite similar. The three body effects are more pronounced for the Dalitz plot distributions to which we turn next.

\begin{figure*}[t]
\centering \includegraphics[keepaspectratio,width=0.46\textwidth]{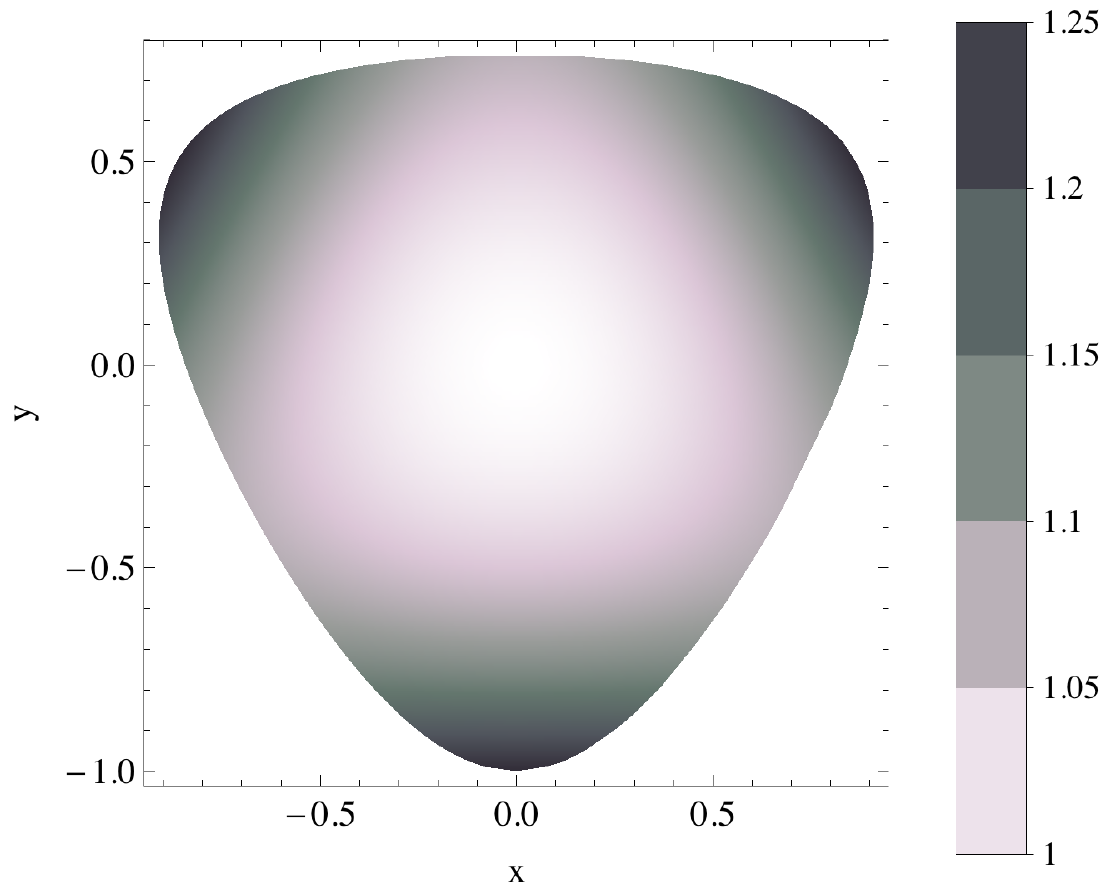}\hspace*{0.02\textwidth}
\includegraphics[keepaspectratio,width=0.46\textwidth]{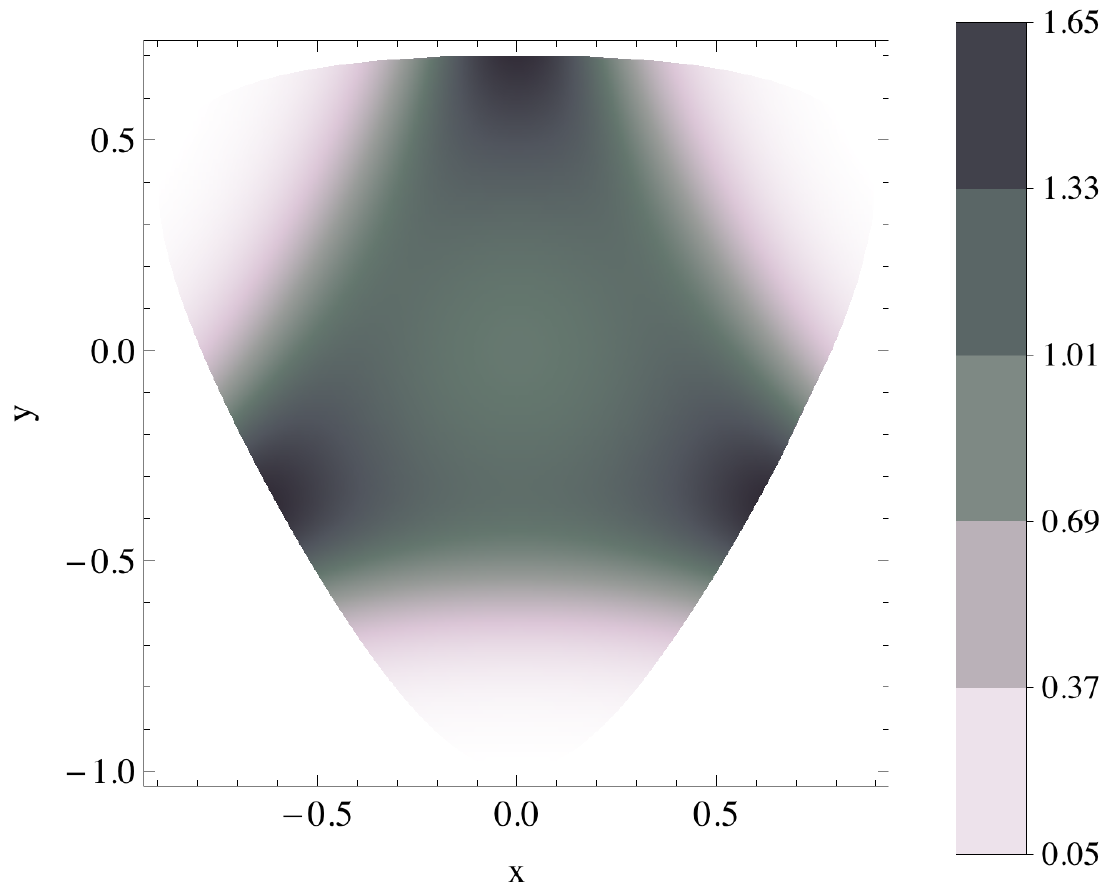}
\caption{\label{Fig:Dalitz} The Dalitz plots for $\omega\rightarrow 3\pi$ (left-hand panel) and $\phi\rightarrow 3\pi$ (right-hand panel) decays. The distributions are divided by the p-wave phase space $P$ and normalized to 1 at $x=y=0$. This is a parameter free result, because we kept only one term in the conformal expansion (\ref{Eq:Sigma}) which is responsible for the overall normalization. See main text for details.}
\end{figure*}

In Fig. \ref{Fig:Dalitz} we show the Dalitz plot distribution in terms of Lorentz-invariant dimensionless parameters,
\begin{eqnarray}\label{Eq:x_y}
x&=&\frac{\sqrt{3}}{Q}(T_1-T_2)=\frac{\sqrt{3}(t-u)}{2M(M-3\,m_\pi)},\nonumber\\
y&=&\frac{3\,T_3}{Q}-1=\frac{3(s_c-s)}{2M(M-3m_\pi)}\,.
\end{eqnarray}
Here $T_i$ is the kinetic energy of the $i$-th pion in the three-particle rest frame and, using the isospin-averaged pion mass,  $Q=M-3m_\pi^2$ and $s_{c}=\frac{1}{3}(M^2+3\,m_\pi^2)$ represents the location of the center of the Mandelstam triangle. Dalitz plot distribution is symmetric under the $x\leftrightarrow -x$ reflection as a consequence of the $t\leftrightarrow u$ symmetry. For $\omega$ decays it is convenient to parametrize the Dalitz plot distribution in terms of a polynomial expansion in $x$ and $y$  around the center of the plot. We follow the procedure outlined in \cite{Niecknig:2012sj} and introduce polar variables
\begin{equation}
x=\sqrt{z}\,\cos\vartheta\,,\quad y=\sqrt{z}\,\sin\vartheta\,,
\end{equation}
and fit the polynomial expansion
\begin{gather}\nonumber
|F_{par}(z,\vartheta)|^2=|N|^2\big(1+2\,\alpha\,z+2\,\beta\,z^{3/2}\sin(3\vartheta)+2\,\gamma\,z^2\\
+2\,\delta\,z^{5/2} \sin(3\vartheta)+\mathcal{O}(z^3)\big)\label{Eq:Fpar}
\end{gather}
to our matrix element. In (\ref{Eq:Fpar}) $N$ is the overall normalization constant. To find Dalitz plot parameters we minimize 
\begin{gather}\nonumber
\bar{\chi}^2=\int\limits_{D}\frac{dz\,d\vartheta}{N_D}\,\left(\frac{P(z,\vartheta)^2(|F_{par}(z,\vartheta)|^2-|F(z,\vartheta)|^2)}{P(0,0)^2|N|^2}\right)^2\\
N_D=\int\limits_{D}dz\,d\vartheta\,,
\end{gather}
where the integration range ($D$) is limited by the Dalitz plot. The results are summarized in Table \ref{Tab:DalitzParameters}. In Table  \ref{Tab:DalitzParameters} we observe a non negligible deviation between the Dalitz plot parameters with and without three body effects. In particular, the three-body effects result in a decrease of intensity by approximately $5\%$ at the boundary of the Dalitz plot and an increase by approximately $2 \%$ in the center. A similar, but even more significant effect is observed for $\phi\rightarrow 3\pi$, where the Dalitz plot intensity decreases at the boundary by $42 \%$ and increases by $6 \%$ in the central  area. In Table \ref{Tab:DalitzParameters} we also compare our results with other theoretical calculations from \cite{Terschlusen:2013iqa} and \cite{Niecknig:2012sj}. We find our Dalitz plot parameters to be quite similar to \cite{Niecknig:2012sj}. We recall that in Ref. \cite{Niecknig:2012sj} the dispersive integral was  extended till infinity, and in order to make that integral convergent, at least one subtraction was required. In our case the unsubtracted dispersive integral is always finite and the number of parameters is determined by (\ref{Eq:Sigma}). From the other side, our results in general smaller than the ones given in \cite{Terschlusen:2013iqa}. The latter calculation was based on a chiral Lagrangian modified by the explicit inclusion of light vector mesons \cite{Terschlusen:2012xw}. In \cite{Terschlusen:2012xw} the unknown coupling constants of the Lagrangian were obtained from the decay properties of the vector mesons \cite{Lutz:2008km}. To this extent, the result of \cite{Terschlusen:2013iqa} provides a good estimate for the decay width, while in the present analysis the decay width was used to fix the normalization. The shortcoming of the approach in \cite{Terschlusen:2013iqa} is that it does not fully comply with unitarity. Though the two-body partial waves were unitarized, the crossed-channel effects were not included. 

\begin{table*}[t]
\centering \caption{\label{Tab:DalitzParameters}Dalitz Plot parameters and $\sqrt{\bar{\chi}^2}$ of the polynomial parametrization (\ref{Eq:Fpar}) for $\omega \rightarrow 3\pi$. In addition to our results we also show the selected results from Niecknig et al. \cite{Niecknig:2012sj} (dispersive study with incorporated crossed-channel effects) and Terschlusen et al. \cite{Terschlusen:2013iqa} (Lagrangian based study with pion-pion rescattering effects).}
\begin{tabular*}{\textwidth}{@{\extracolsep{\fill}}llllll@{}} 
\hline\hline   & $\alpha\times 10^{3}$ & $\beta\times 10^{3}$ & $\gamma\times 10^{3}$ & $\delta\times 10^{3}$ & $\sqrt{\bar{\chi}^2}\times 10^{3}$
\\ \hline
This paper  ($\hat F=0$)                           & 136     & - & - & - & 3.5\\
This paper (full)                                         & 94       & - & - & - & 3.2\\
Niecknig et al. \cite{Niecknig:2012sj}        & 84...96 & - & - & - & 0.9...1.1\\
Terschlusen et al. \cite{Terschlusen:2013iqa} & 202 & - & - & - & 6.6\\
\hline
This paper  ($\hat F=0$)                           & 125      & 30      & - & - & 0.74\\
This paper (full)                                               & 84        & 28      & - & - & 0.35\\
Niecknig et al. \cite{Niecknig:2012sj}        & 74...84 & 24...28 & - & - & 0.052...0.078\\
Terschlusen et al. \cite{Terschlusen:2013iqa} & 190 & 54 & - & - & 2.1\\
\hline
This paper  ($\hat F=0$)                           &  113    &  27       & 24     & - & 0.1\\
This paper (full)                                               & 80       & 27        & 8      & -  & 0.24\\
Niecknig et al. \cite{Niecknig:2012sj}        & 73...81 & 24...28 & 3...6 & - & 0.038...0.047\\
Terschlusen et al. \cite{Terschlusen:2013iqa} & 172 & 43 & 50 & - & 0.4\\
\hline
This paper  ($\hat F=0$)                           & 114 & 24 & 20 & 6 & 0.005\\
This paper (full)                                                & 83 & 22 & 1 & 14 & 0.079\\
Niecknig et al. \cite{Niecknig:2012sj}        & 74...83 & 21...24 & 0...2 & 7...8& 0.012...0.011\\
Terschlusen et al. \cite{Terschlusen:2013iqa} & 174 & 35 & 43 & 20 & 0.1\\
\hline\hline
\end{tabular*}
\end{table*}

On the experimental side, the situation as follows. The measurements of $\phi \rightarrow 3\pi$ were performed by KLOE \cite{oai:arXiv.org:hep-ex/0303016} and CMD-2 \cite{Akhmetshin:2006sc} collaborations. As for $\omega$ decay we expect new data from CLAS12, WASA at COSY and KLOE collaborations. Since the main purpose of the present paper is to outline a novel theoretical scheme, we postpone a comprehensive data analysis to the future and for now only consider the application to EM transition form factors of $\omega/\phi$. In particular, the transition $\omega \rightarrow \pi \gamma^*$ is of interest since the existing data in the time-like region seems to be incompatible with the vector meson dominance model (VMD) \cite{Arnaldi:2009aa, Usai:2011zza}.

\subsection{$\omega/\phi \rightarrow \pi \gamma^*$}
\label{Sec:em}

In this section we discuss the EM transition form factors of the $\omega$ and $\phi$ mesons. The Dalitz decay of the vector mesons into pion and a lepton pair
\begin{multline}
\langle \pi^0(p_0)\,l^+(p_+)\, l^-(p_-)\,|\,T\,|\,V(p_V,\lambda)\rangle=\\ (2\pi)^4\,\delta(p_V-p_0-p_+-p_-)\,H_{V\pi}\,,
\end{multline}
can be described by the following amplitude \cite{Landsberg:1986fd}
\begin{multline}\label{Eq:EMFF_covariant_form}
H_{V\pi}=  \epsilon^{\mu}(p_V,\lambda) f_{V\pi}(s)\,\epsilon_{\mu\nu\alpha\beta}\,p_0^\nu\,q^\alpha\\
\frac{i e^2}{s}\,\bar{u}(p_-,\lambda_-)\,\gamma^\beta\,\upsilon(p_+,\lambda_+),
\end{multline}
which describes the product of the hadronic current, the photon propagator and the lepton current. In addition to a kinematical factor, the hadron current is given in terms of the form factor $f_{V\pi}(s)$. In (\ref{Eq:EMFF_covariant_form}) $q$ is the momentum of the virtual photon with invariant mass $s=q^2=(p_++p_-)^2$ and $\bar{u}$, $\upsilon$ stand for Dirac spinors of the two leptons. The single differential decay rate normalized by the real photon decay width is given by $\Gamma_{V\rightarrow \pi\gamma}$, and can be written as
\begin{multline}\label{Eq:Single_differential_decay_rate}
\frac{1}{\Gamma_{V\rightarrow \pi\gamma}}\frac{d\Gamma}{ds}=\frac{e^2}{12\,\pi^2}\,|F_{V\pi}(s)|^2\,\sqrt{1-\frac{4\,m_l^2}{s}}\left(1+\frac{2\,m_l^2}{s}\right)\\
\frac{1}{s}\left[\left(1+\frac{s}{M^2-m^2}\right)^2-\frac{4\,M^2s}{(M^2-m^2)^2}\right]^{3/2}
\end{multline}
where $e=0.303 = \sqrt{4\,\pi\,\alpha_{em}}$ is the electric charge, $m_l$ is the lepton mass,
\begin{equation}
\Gamma_{V\rightarrow \pi\gamma}=\frac{e^2\,(M^2-m_\pi^2)^3}{96\pi\,M^3}\,|f_{V\pi}(0)|^2\,,
\end{equation}
and $F_{V\pi}(s)$ is the hadronic form factor normalized to unity at the photon point $s=0$,
\begin{equation}
F_{V\pi}(s)=\frac{f_{V\pi}(s)}{f_{V\pi}(0)}\,.
\end{equation}

\begin{figure}[t]
\centering
\includegraphics[keepaspectratio,width=0.4\textwidth]{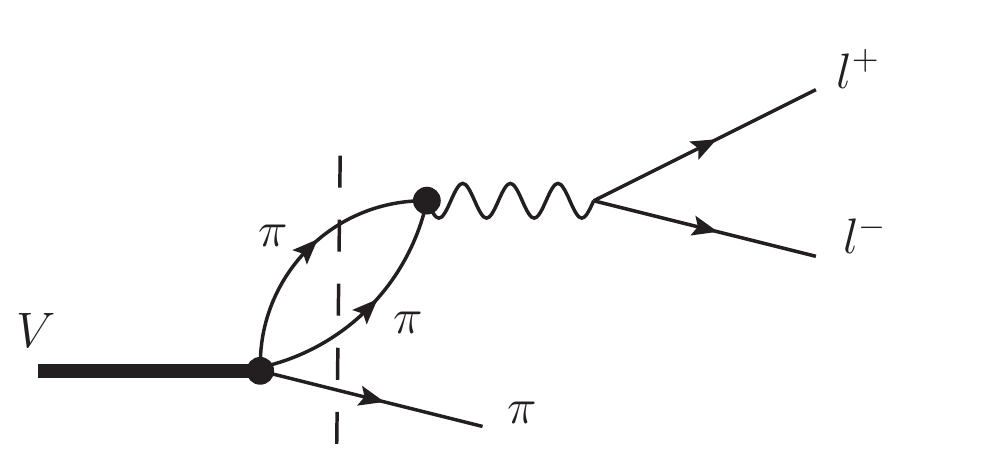}
\caption{\label{Fig:Diagram_EMFF}Schematic representation of the discontinuity for the electromagnetic transition form factor.}
\end{figure}

\begin{figure*}[t]
\center{\includegraphics[keepaspectratio,width=0.96\textwidth]{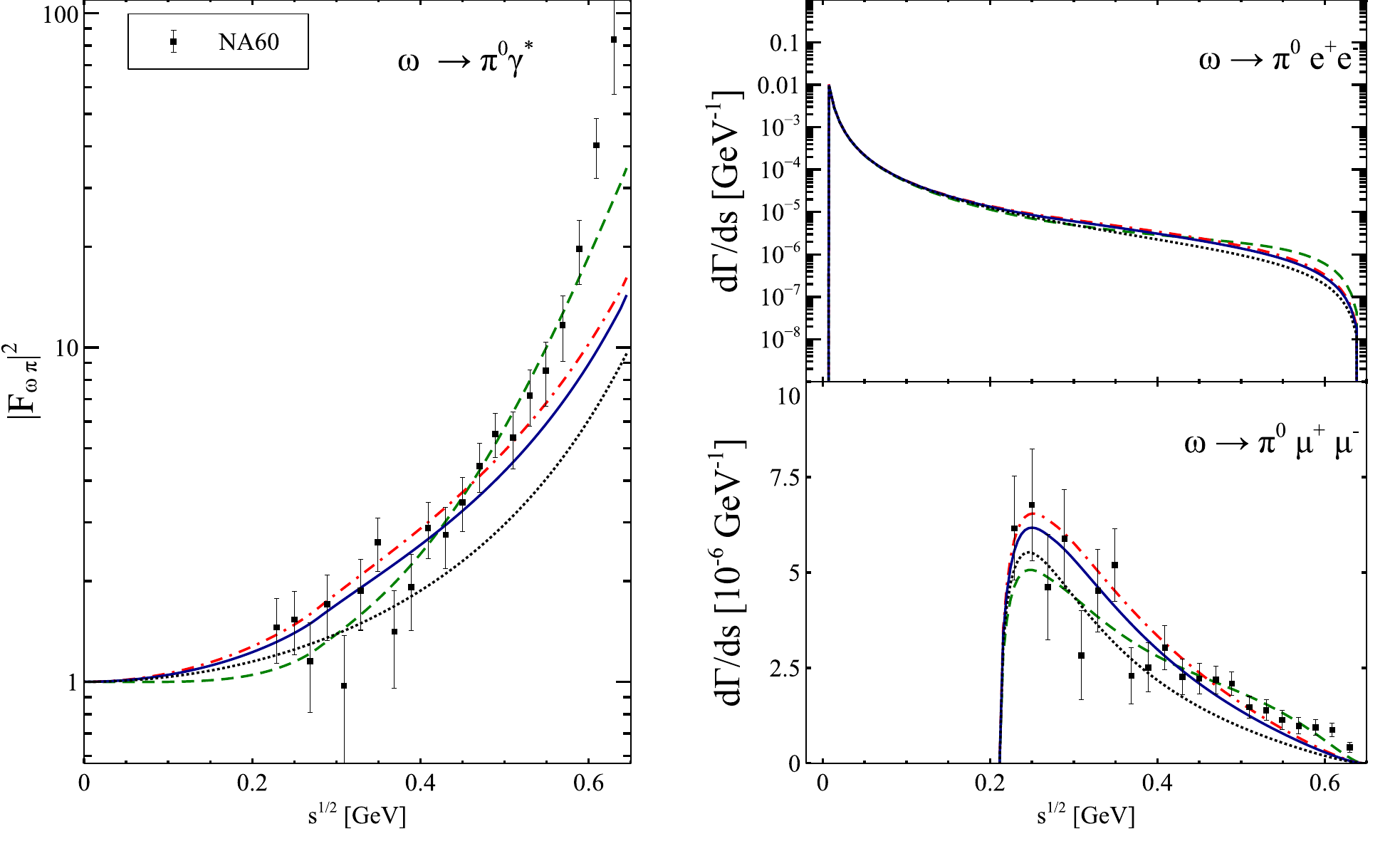}}
\caption{\label{Fig:EMFF_omega}The Electromagnetic form factor for $\omega\rightarrow \pi^0\gamma^*$ (left panel), the differential decay rate $\omega \rightarrow \pi^0e^+e^-$ (top right) and  the differential decay rate $\omega \rightarrow \pi^0e^+e^-$ (bottom right). Data for the form factor is taken from \cite{Arnaldi:2009aa}, while for the single-differential decay rate were calculated using Eq.(\ref{Eq:Single_differential_decay_rate}). The dotted line is the VMD model (\ref{Eq:VMD}), while the solid, dash-dotted and dashed lines correspond to a truncation in the expansion (\ref{Eq:Disp_relation_EMFF}) at order $0$, $1$, $2$ respectively.}
\end{figure*}
\begin{figure*}[t]
\center{\includegraphics[keepaspectratio,width=0.96\textwidth]{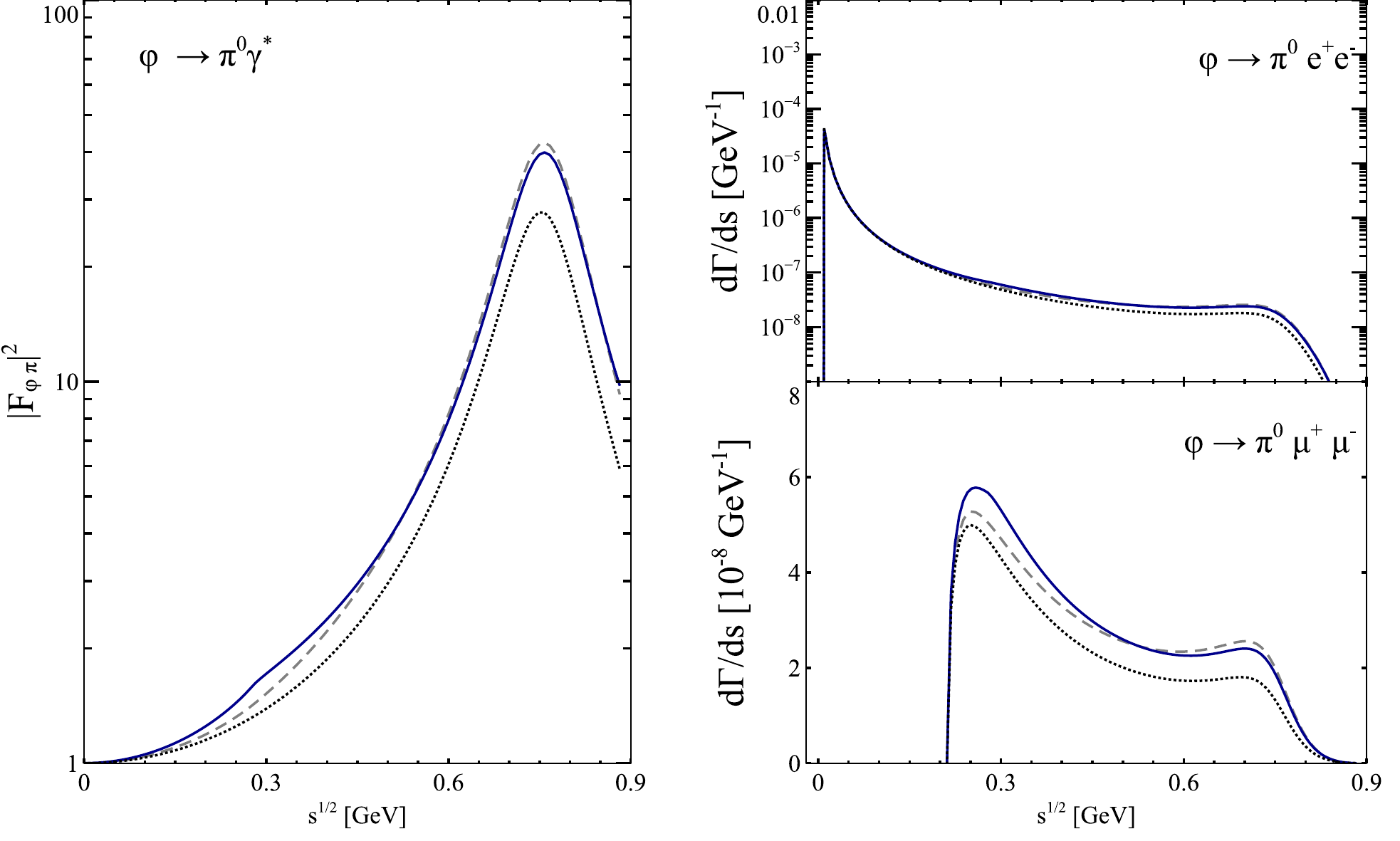}}
\caption{\label{Fig:EMFF_phi}The Electromagnetic form factor for $\phi\rightarrow \pi^0\gamma^*$ (left panel), the differential decay rate $\phi \rightarrow \pi^0e^+e^-$ (top right) and  the differential decay rate $\phi \rightarrow \pi^0e^+e^-$ (bottom right). The dotted line is the VMD approach (\ref{Eq:VMD}), the solid line corresponds to a truncation in the expansion (\ref{Eq:Disp_relation_EMFF}) at $0^{th}$ order and the dashed line is the same as the solid line but without three body effects.}
\end{figure*}
In the elastic approximation, illustrated in Fig.~\ref{Fig:Diagram_EMFF} the discontinuity of the EM transition form factors across the $\pi\pi$ cut \cite{Koepp:1974da} is proportional to the $V\rightarrow 3\pi$ decay amplitude (see Eq. (\ref{Eq:covariant_form})) and the pion vector form factor $F_{\pi}(s)$, 
\begin{equation}
\text{Disc}\,f_{V\pi}(s)=\frac{\rho^3(s)\,s}{128\,\pi}\,F_{\pi}^*(s)\int_{-1}^{1}dz'(1-z'^2)\,F(s,t',u')
\end{equation}
The dispersion relation for the form factor can therefore be written as, 
\begin{eqnarray}
f_{V\pi}(s)&=&\int_{s_\pi}^{s_i}\frac{ds'}{\pi}\frac{\text{Disc}\,f_{V\pi}(s')}{s'-s}+\tilde{\Sigma}(s)\nonumber\\
\tilde{\Sigma}(s)&=&\sum_{i=0}^{\infty}b_i\,\omega^i(s)\label{Eq:Disp_relation_EMFF}
\end{eqnarray}
where we separated the elastic and inelastic contributions in a similar fashion as for the $V\rightarrow3\pi$ amplitude.  The inelastic contribution is defined by a map of the $s$-plane cut above $s=s_i$ and is thus given by the same function $\omega(s)$, {\it cf.}  Eq.~(\ref{Eq:Conformal_mapping_function}). However, the coefficients, $b_i$, specify the form factor and are different from those in Eq.~(\ref{Eq:Sigma}). We remove the unphysical discontinuity at  $s=s_i$ using the procedure outlined in the previous section, {\it cf.} Eq.~(\ref{Eq:Regularization}). As for the pion vector form factor, we employ the parametrization that was used by the Belle Collaboration \cite{Fujikawa:2008ma}, which we refer to as $F^{Belle}(s)$. However, in order to satisfy the Watson theorem \cite{Watson:1954uc}, we modify $F^{Belle}(s)$ and for the vector form factor use $F_\pi(s) =|F^{Belle}|\exp(i\delta(s))$, where $\delta$ is the $\pi\pi$ P-wave phase shift taken from \cite{GarciaMartin:2011cn}. We have checked that the effect of this modification on the description of the experimental data is negligible in the energy range $s=[s_\pi,\,s_i]$. We stress that thanks to the separation of the elastic and inelastic contributions of $f_{V\pi}(s)$ there is no need for additional assumptions regarding the behavior of the $\pi\pi$ phase shift beyond the elastic region, in contrast to \cite{Schneider:2012ez}.

Recently the NA60 Collaboration \cite{Arnaldi:2009aa, Usai:2011zza} reported a new measurement of the electromagnetic transition form factor from the decay $\omega\rightarrow \pi^0 \mu^+\mu^-$. This process is interesting, because the most common approach, namely VMD \cite{Sakurai1969}, 
\begin{eqnarray}\label{Eq:VMD}
F_{\text{VMD}}(s)=\frac{m_\rho^2}{m_\rho^2-s-i\,\sqrt{s}\,\Gamma(s)}
\end{eqnarray}
dramatically fails to reproduce the data. Note, that in the case of $\phi \rightarrow \pi \gamma^*$ the rho meson pole occurs in the physical region, and therefore we included the width in the denominator of (\ref{Eq:VMD}),
\begin{gather}\label{Eq:VMD2}
\Gamma(s)=\Gamma_\rho\left(\frac{p_\pi(s)}{p_\pi(m_\rho^2)}\right)^3\frac{m_\rho^2}{s}\,,
\end{gather}
where $p_\pi$ is the pion momentum in the rho meson center of mass and $\Gamma_\rho=150$ MeV. This change is not important for $\omega \rightarrow \pi \gamma^*$ decay, where the narrow width approximation works very well.

In the following, we will compare our results with VMD (\ref{Eq:VMD}), Schneider et al. \cite{Schneider:2012ez} and Terschlusen et al. \cite{Terschlusen:2012xw}. In \cite{Schneider:2012ez} the dispersive analysis of three pion decays of $\omega$ and $\phi$ mesons was extended to EM transition form factors. Similarly to $\omega/\phi \rightarrow 3\pi$ analyses, the dispersive integral was extended to infinite energies and inelastic contributions were suppressed by subtractions. The $\pi\pi$ p-wave phase shift was assumed to have asymptotic behavior of $\delta(s\rightarrow\infty)=\pi$. In the analysis of \cite{Terschlusen:2012xw} the chiral Lagrangian with vector mesons \cite{Terschlusen:2012xw,Terschlusen:2013iqa, Leupold:2008bp} was used.

The $\omega\rightarrow \pi^0\gamma^*$ EM transition form factors is shown in Fig. \ref{Fig:EMFF_omega}  together with the  differential $\omega\rightarrow \pi^0 e^+e^-$ and $\omega\rightarrow \pi^0 \mu^+\mu^-$ decay rates. The various lines illustrate the effect of higher order terms in the expansion of the inelastic contribution in terms of $\omega(s)$ (\ref{Eq:Disp_relation_EMFF}).  The $b_0=-0.194$ parameter is determined by comparing with the real-photon decay width $\Gamma^{exp}_{\omega\rightarrow\pi^0\gamma}=0.703$ MeV \cite{PDG-2012}, while the other $b_{i\geq1}$ parameters were obtained from the fitting the EM form factor data. As it can be seen in Fig. \ref{Fig:EMFF_omega}, keeping only one term in the conformal expansion already gives a reasonable description. It improves the slope of the VMD curve towards the data with $\chi^2/\text{d.o.f.}\backsimeq2.5$ compared to $\chi^2/\text{d.o.f.}\backsimeq4.6$ using the VMD model. The quality of the data description is similar to that of \cite{Schneider:2012ez} and somewhat worse when compared to  \cite{Terschlusen:2012xw} which corresponds to $\chi^2/\text{d.o.f.}\backsimeq1.8$.  In Fig. \ref{Fig:EMFF_omega} we also show the single-differential decay rates of $\omega\rightarrow \pi^0 e^+e^-$ and $\omega\rightarrow \pi^0 \mu^+\mu^-$. The kinematic factors suppress the large invariant mass region and therefore the branching ratios agree very well with the experimental values \cite{PDG-2012}, 
\begin{eqnarray}
\mathcal{B}^{th}(\omega\rightarrow \pi^0 e^+ e^-)&=&7.8\cdot 10^{-4}\nonumber\\
\mathcal{B}^{exp}(\omega\rightarrow \pi^0 e^+ e^-)&=&(7.7\pm 0.6)\cdot 10^{-4}
\end{eqnarray}
and
\begin{eqnarray}
\mathcal{B}^{th}(\omega\rightarrow \pi^0\mu^+\mu^-)&=&0.96\cdot 10^{-4}\nonumber\\
\mathcal{B}^{exp}(\omega\rightarrow \pi^0\mu^+\mu^-)&=&(1.3\pm 0.4)\cdot 10^{-4}\,.
\end{eqnarray}

Since the experimental data are not very precise, we decided to estimate the coefficients $b_i$ of (\ref{Eq:Disp_relation_EMFF}) by matching our amplitude to $\chi$PT with vector mesons \cite{Terschlusen:2012xw} at $s=0$. We remark, that the expansion coefficients $b_i$ can be uniquely determined by the first $i$ derivatives of $\tilde{\Sigma}(s)$ at the expansion point $s=s_E=0$ (see Eq. (\ref{Eq:Conformal_mapping_function})). 
We find the following results
$$
\begin{tabular*}{\columnwidth}{@{\extracolsep{\fill}}llll}
\hline   & $b_0$ & $b_1$ & $\chi^2/\text{d.o.f.}$\\
\hline
Data fit (only $b_1$) & -0.194& 4.96 & 2.4 \\
Matching to $\chi$PT with VM  & -0.148  & 9.33 & 2.4 \\
\hline
\end{tabular*}
$$
which improve but do not resolve the disagreement between the data and our description for the last few data points (dot-dashed curve in Fig. \ref{Fig:EMFF_omega}). As a phenomenological test we decided to add one more term in the conformal expansion and all together fit $b_1$ and $b_2$ to the NA60 data (dashed curve in Fig. \ref{Fig:EMFF_omega}). The resulting parameters are $b_1=-23.7$ and $b_2=484.4$ with $\chi^2/\text{d.o.f.}\backsimeq1.3$. The fit indicates a significant change in the parameter $b_1$ (even different sign). The variation of fit parameters is consistent with the strong rise of the form factor, which is modeled, through $\omega(s)$, by a singularity at the inelastic threshold. It is doubtful, however, that this would be the correct explanation. An independent measurement of the $\omega$ and $\phi$ form factors should help resolve this puzzle. 
     
In the elastic region the discontinuity is exact up to uncertainties in the pion-pion amplitude. For the inelastic region we use a parametrization that we fit to the data. One can contemplate a study of the theoretical uncertainties, for example,  by using different forms of the conformal mapping. We checked that changing the expansion point in Eq. (\ref{Eq:Conformal_mapping_function}) to $s=4m_\pi^2$ produces negligible effects. Another question pertains to  the criteria for choosing the number of terms in the  expansion and possible constraints on the conformal coefficients. Since we are seeking a description of the data in the decay region, the energies are always smaller than $s_i$,  which should  guarantee   good convergence of the conformal expansion. For example, at the origin $\omega(s=0)=0$, and at the edges of the decay region $\omega(s=(M_\omega-m_\pi)^2)=0.133$ and $\omega(s=(M_\phi-m_\pi)^2)=0.358$. Therefore, a few terms in the expansion produce reasonable results. The possible estimations on the size of the  conformal coefficients can come, for example, from  $\chi$PT at low energies or other phenomenological analyses, {\it e.g.} that include explicit coupled-channels. We find, for example, that in the fit with three terms in conformal expansion the values of the parameters are consistent  with the results of  \cite{Ananthanarayan:2014pta}.
     
Figure \ref{Fig:EMFF_phi} shows the results for the $\phi$ meson decays. Since there are no experimental measurements, we keep only one term in the conformal expansion (\ref{Eq:Disp_relation_EMFF}) which is fixed by the experimental real-photon decay width, $\Gamma^{exp}_{\phi\rightarrow\pi^0\gamma}=5.41$ keV \cite{PDG-2012}. For the branching ratio, it then leads to

\begin{equation}
\mathcal{B}^{th}(\phi\rightarrow \pi^0 e^+ e^-)=1.45\cdot 10^{-5}\,,
\end{equation} 
which compares favorably with the experimental value \cite{PDG-2012} of, 
\begin{equation} 
\mathcal{B}^{exp}(\phi\rightarrow \pi^0 e^+ e^-)=(1.12\pm 0.28)\cdot 10^{-5}\,.
\end{equation}  
The predicted branching ratio to muons is 
\begin{eqnarray}
\mathcal{B}^{th}(\phi\rightarrow \pi^0 \mu^+ \mu^-)&=&3.9\cdot 10^{-6}\,.
\end{eqnarray}
Finally in Fig. \ref{Fig:EMFF_phi} we show the sensitivity of the $\phi$ form factor to the three-body effects in $\omega \to 3\pi$ decay.  We confirm the findings of \cite{Schneider:2012ez}, that there is an enhancement at the two-pion threshold due to cross-channel re-scattering effects. As another theoretical study we refer \cite{Pacetti:2009pg}, where the EM form factor in the resonance region was parametrized by a sum of the vector propagators weighted by the corresponding coupling constants.

We emphasize that our approach is restricted to low energies. It can however be matched onto a particular high energy model, by imposing additional constraints on the coefficients of conformal mapping.

\section{Conclusions}
\label{Sec:Conclusion}

In this paper we have analyzed three-pion decays and electromagnetic form factors of $\omega/\phi$ within a dispersive formalism that is based on the isobar decomposition and sub-energy unitarity. The important input is the P-wave $\pi\pi$ scattering amplitude that is available from \cite{GarciaMartin:2011cn}. By means of the dispersion relation we separated the contribution from the elastic and inelastic channels. The latter was parametrized by a series in a suitable conformal variable and the coefficients of this expansion play the role of the subtraction constants.  When the partial wave expansion is truncated, constraints from Regge theory on the high energy behavior are missing. In this case partial wave dispersion relations do not have unique solutions as they depend on the assumed asymptotic behavior. We have presented an alternative method for incorporating three-body effects that alleviates some of the deficiencies  when dealing with inelastic contributions to partial waves dispersion relations. The unknowns are parametrized though a conformal expansion with coefficients that can either  be  fitted to the data or determined by comparing with other theoretical studies, {\it e.g.} Lattice QCD of EFT expansion. To  properly incorporate the high-energy behavior, however, it is necessary to build in aspects of the Regge theory which we leave for future investigations.

We presented the single-differential and Dalitz plot distributions, where we found non-negligible three body effects. We also found our results to be similar to those of Ref. \cite{Niecknig:2012sj}, where a standard subtraction procedure was applied. As a straightforward application of the three-body amplitude we studied electromagnetic form factors for $\omega/\phi$ mesons. The results improve over the simple VMD model, however, our theoretical analysis and the other studies \cite{Schneider:2012ez,Terschlusen:2012xw} predict the EM transition form factor for $\omega \rightarrow \pi \gamma^*$ to be smaller at $s=(M_{\omega}-m_\pi)^2$ than that measured by the NA60 Collaboration. To shed more light on the intrinsic dynamics of hadrons at low energies the experimental analysis of the Okubo-Zweig-Iizuka-suppressed decay $\phi \rightarrow \pi^0 l^+ l^-$ is highly desirable. The shape of the latter is predicted within our framework.

As a next step we plan to perform the data analysis of the upcoming $\omega\rightarrow 3\pi$ JLab g12 data. Note, that the same method can be applied to treat $D$ and $B$ mesons three body decays. Another prospect is the hadronic light-by-light contribution to the anomalous magnetic moment of the muon \cite{Colangelo:2014pva}, where $\omega/\phi \rightarrow \pi \gamma^*$ serve as input ingredients to the pion transition form factor $F_{\pi^0\gamma^*\gamma^*}$ and $\gamma^*\gamma^*\rightarrow \pi\pi$ partial waves.

All the material, including the codes are available in interactive form online~\cite{website}.

\begin{acknowledgments}
We are grateful for useful discussions with M.R. Pennington, V. Mokeev and D.J. Wilson.
This material is based upon work supported in part by the U.S. Department of Energy, Office of Science, Office of Nuclear Physics under contract DE-AC05-06OR23177.
This work was also supported in part by the U.S. Department of Energy under Grant No. DE-FG0287ER40365, National Science Foundation under Grant PHY-1415459.  

\end{acknowledgments}

\bibliographystyle{prsty}
\bibliography{VtoPPP}

\end{document}